\documentclass[journal]{IEEEtran}

\usepackage{flushend}
\usepackage{cite}
\usepackage[font=small,labelfont=bf]{caption}
\usepackage{subcaption}
\def\BibTeX{{\rm B\kern-.05em{\sc i\kern-.025em b}\kern-.08em
    T\kern-.1667em\lower.7ex\hbox{E}\kern-.125emX}}
    
\usepackage{amsmath,amssymb,amsfonts,bm,amsthm}
\usepackage{algorithmic}
\usepackage[ruled,vlined]{algorithm2e}
\usepackage{siunitx}
\usepackage{tcolorbox}
\usepackage{pgfplots}
\usepackage{grffile}
\pgfplotsset{compat=newest}
\usetikzlibrary{math,calc,plotmarks,arrows.meta}
\usetikzlibrary{arrows.meta}
\usepgfplotslibrary{patchplots}
\usepackage{graphicx}
\usepackage{xcolor}
\usepackage{float}
\usepackage[hidelinks]{hyperref}
\usepackage{cleveref}
\usepackage{colortbl}
\usepackage{tikz}
\usepackage{multirow}

\DeclareMathOperator*{\p}{\mathcal{P}}

\renewcommand{\Vec}{\mathbf}
\definecolor{navyblue}{rgb}{0.00000,0.44700,0.74100}%

\SetKwInput{KwInput}{Input}                
\SetKwInput{KwOutput}{Output}              

\begin{document}
	\title{Why Performance Metrics Overpromise in Auditory Attention Decoding: an Information-Theoretic Reappraisal}
	
	\author{Nicolas~Heintz, Simon~Geirnaert, Tom~Francart, and~Alexander~Bertrand,~\IEEEmembership{Senior~Member,~IEEE}%
	\thanks{N. Heintz, S. Geirnaert and A. Bertrand are with KU Leuven, Department of Electrical Engineering (ESAT), STADIUS Center for Dynamical Systems, Signal Processing and Data Analytics and also with Leuven.AI - KU Leuven institute for AI, Kasteelpark Arenberg 10, B-3001 Leuven, Belgium (e-mail: nicolas.heintz@esat.kuleuven.be, simon.geirnaert@esat.kuleuven.be, alexander.bertrand@esat.kuleuven.be). 

    T. Francart is with KU Leuven, Department of Neurosciences, Research Group ExpORL, Herestraat 49 box 721, B-3000 Leuven, Belgium (e-mail: tom.francart@kuleuven.be).

    The authors acknowledge the financial support of the FWO (Research Foundation Flanders) for project 1S31524N, G026026N, 1242524N, and G081722N, the Flemish Government (AI Research Program), and Internal Funds KU Leuven (project C3/25/017, IDN/23/006 and C14/25/108. For the purpose of open access, the author has applied a CC BY public copyright licence to any Author Accepted Manuscript version arising from this submission.}}

	\maketitle

	\begin{abstract}
         Auditory attention decoding (AAD) algorithms are predominantly evaluated in a steady state where a listener continuously attends to the same speaker, using metrics such as accuracy and information transfer rate. However, such metrics fail to account for the (in-)dependence of an AAD prediction with respect to previous predictions. In this paper, we argue that failing to take this dependence into account in the algorithm evaluation can lead to severe misrepresentations of the true performance of an AAD algorithm. 
         
         We therefore introduce the relative Incremental Mutual Information (rIMI); the rate at which a new prediction removes the remaining uncertainty about the identity of the attended speaker. This allows us to track how much new, useful information a prediction actually generates on top of the information already obtained from previous predictions. By investigating the rIMI and the behaviour of AAD models around attention switches, we demonstrate that recent direct-classification AAD algorithms are not superior to traditional AAD algorithms based on stimulus reconstruction, despite what accuracy alone may suggest. We also demonstrate how these direct-classification AAD predictions are severely influenced by irrelevant feature drifts, which artificially inflates accuracies by leaking information across windows, and even across trials.
	\end{abstract} 
	
	\section{Introduction}
    When multiple people are talking simultaneously, our brain can specifically attend to a single speaker and ignore unattended speech \cite{Cherry1953}. This allows us to understand speech, even in extremely noisy scenarios. However, people suffering from hearing loss quickly lose this capability \cite{Shinn-Cunningham2008,presaccoSpeechinnoiseRepresentationAging2019}. While current hearing aids are increasingly effective at enhancing target speech, they often fail to identify the attended speaker in complex listening environments. A promising solution is to decode the attended speaker from the brain activity, for example based on non-invasive electroencephalography (EEG). This is known as auditory attention decoding (AAD) \cite{OSullivan2014, DeCheveigne2018, Geirnaert2020a}. 

    AAD algorithms are typically evaluated based on two key metrics: their time resolution and their decoding accuracy. In general, the resolution of an AAD algorithm is estimated based on the algorithm's decision window length, i.e. the amount of (EEG) data based on which a decision is made. 
    
    Traditionally, the auditory attention was decoded using stimulus reconstruction algorithms where an EEG decoder is trained to reconstruct the attended speech envelope from the neural responses in the EEG signals, either with linear or non-linear models. The attended speaker is then identified by computing the correlation between the decoded EEG (i.e., the reconstructed stimulus) and the speech envelope of each speaker. Due to the low SNR of this process, stimulus reconstruction algorithms typically require up to \SI{10}{\second} to generate a sufficiently accurate prediction of the attended speaker \cite{OSullivan2014, DeCheveigne2018, nguyenAADNetEndtoEndDeep2025}. However, recent AAD research shifts more and more towards spatial or direct-classification algorithms, where the identity of the attended speaker or the locus of attention is directly decoded from EEG without explicitly correlating it to the speech signals \cite{Geirnaert2020CSP, geirnaertRiemannianGeometryBasedDecoding2021a, Zhang, suSTAnetSpatiotemporalAttention2022, sridharImprovingAuditoryAttention2025, nguyenAADNetEndtoEndDeep2025}. Although such methods currently suffer from poor generalisation, and regularly overfit on irrelevant confounds such as eye gaze \cite{rotaruWhatAreWe2024,ivucicImpactCrossValidationSchemes2024,yanOverestimatedPerformanceAuditory2025}, there remains a strong focus on such approaches within the AAD literature. This focus is driven by their seemingly superior potential to achieve very high accuracies on very short window lengths (\SIrange{0.1}{2}{\second}).

    However, we show in this paper that this potential of direct-classification AAD algorithms to outperform stimulus reconstruction AAD algorithms is largely a mirage, driven by incomplete metrics that mask the true performance of an AAD algorithm. At the core of this mirage is the fact that consecutive direct-classification AAD predictions consistently demonstrate a high conditional self-dependence. This means that a large portion of the variance in the AAD predictions can be explained by previous predictions. We will demonstrate that, when the conditional self-dependence is not accounted for, the performance of an algorithm can be severely overestimated. We argue that this overestimation is further driven by the tendency to evaluate accuracy in sustained-attention experiments, where attention switches are absent or infrequent within an experiment trial. We show that, due to these temporal dependencies, accuracy measured under sustained-attention conditions is a poor predictor of performance in the vicinity of an attention switch. 

    In this paper, we use information theory to study how ignoring the conditional self-dependence can lead to a misleading picture of the performance of an algorithm. We also propose a new metric, which allows us to separate repeated information in consecutive AAD predictions from actual, novel information. This aids us to study the true performance of several AAD algorithms, even in datasets where there are no or limited attention switches to evaluate the true performance at these switch events. 

   We explore in depth where the conditional self-dependence comes from, how it can affect performance, and how we can correct for it. However, most important is how the performance overestimations can be avoided in the future. We therefore propose a set of concrete guidelines for the evaluation of AAD algorithms, which are summarised below and further justified throughout the paper. While these guidelines emerge naturally from the information-theoretic analysis presented here, we highlight them already in the introduction, as we believe they constitute the paper's most important contribution. 

    In Section \ref{sec:MI-TemporalDependence}, we clarify in more detail how ignoring the conditional self-dependence can lead to overestimated performances. In Section \ref{sec:informationTheory}, we formulate the information-theoretic metrics that can correct this overestimation. In Section \ref{sec:MI-MutualInformation}, the core methods to estimate the information-theoretic metrics are explained. We further detail how all experiments were performed in Section \ref{sec:MI-Experiments}. Finally, the results are displayed and discussed in Section \ref{sec:MI-Results}.

    \begin{tcolorbox}[colback=navyblue!5!white,colframe=navyblue!75!black,title=How to report AAD performances fairly]
        \begin{itemize}
        \item \textbf{Create a buffer between training and test data.} \newline
            Supervised training data and test data should be separated by as much time as possible. At least five minutes, but ideally over more than one test session. A LOTO cross-validation is often not enough, as temporal dependencies can sometimes persist across trials. 
            \item \textbf{Validate around attention switches.} \newline
            An AAD model should always be tested on data with attention switches (as many as feasible). The performance in steady state and around an attention switch should be documented separately.
            \item \textbf{Avoid confounds in the data.} \newline
            Avoid data where confounds such as eye gaze or muscle artifacts can be abused to infer the attention state. Such artifacts cannot be completely removed with preprocessing. 
            \item \textbf{Measure the rate of information gain.} \newline
            The relative Incremental Mutual Information (rIMI) metric proposed in this paper can optionally be used to compare AAD algorithms, in particular when no or limited attention switches are present in the data.  It estimates what fraction of the remaining uncertainty about the current attention state a prediction removes, addressing the shortcomings of traditional accuracy metrics when measured on sustained attention data. 
        \end{itemize}
    \end{tcolorbox}
    
    \section{How Temporal Dependence inflates Performance}
    \label{sec:MI-TemporalDependence}
	At first glance, it may not be immediately evident why conditional self-dependence matters. \textit{``If an algorithm can predict the identity of the attended speaker with high accuracy using a single short window, why would it matter if the prediction in the next window is conditionally dependent?"} To aid with this intuition, we identify three mechanisms where the performance of an algorithm can be incorrectly inflated by ignoring the conditional self-dependence, depending on the underlying cause of the dependencies. 

    \subsection{An inflated temporal resolution}
    The first case is the most straightforward, and happens when the algorithm's AAD prediction scores do not change fast enough after an attention switch to be significantly different in consecutive windows. When the listener switches attention, it could take several decision windows before the prediction has fully flipped from the first to the second attended speaker. Since it would take more than 1 window to detect an attention switch, the temporal resolution is lower than the window length would suggest, even if you reach near-perfect accuracy in steady state on a single window. This is comparable to an oversampled signal that is sampled above the Nyquist rate.

    A typical example of inflated temporal resolutions is when the auditory attention is predicted on \SI{0.1}{\second} windows, as often found in recent AAD papers \cite{Zhang, suSTAnetSpatiotemporalAttention2022}. Even if you could reliably decode the auditory attention using these short windows in a steady state, it is unlikely that you could accurately track constant sub-second attention switches (unless explicitly proven otherwise). The brain cannot switch attention in such a short amount of time. Moreover, a \SI{0.1}{\second} resolution is impossible by construction, as the bandpass filters that are present in almost all preprocessing pipelines already smear information over multiple \SI{0.1}{\second} windows. 
    
    Another example is the postprocessing that is typically done to smoothen AAD predictions before applying it to a gain control algorithm \cite{Presacco2019, hjortkjaerRealtimeControlHearing2025, Geirnaert2020a,heintzProbabilisticGainControl2024, heintzPostprocessingEEGbasedAuditory2025}. Although these algorithms typically start from short windows (e.g., \SI{1}{\second}), they integrate multiple decisions over time such that it takes much longer to actually detect an attention switch \cite{heintzProbabilisticGainControl2024, heintzPostprocessingEEGbasedAuditory2025}. It would thus be unfair to merely report the accuracy of such postprocessing algorithms and compare them to the accuracies of other AAD algorithms on \SI{1}{\second} windows. 

    \subsection{Temporal drift as a confounding feature}
    Conditionally dependent consecutive predictions are often caused by a temporal dependence in the feature space. This means that feature vectors observed in close succession, are also expected to be close to each other in the feature space. This temporal proximity can be driven by processes that are completely irrelevant to the attention process, such as slowly changing statistics of EEG due to the (background) neural processes, akin to Brownian motion \cite{xuBewareOverestimatedDecoding2024}.

    Since AAD algorithms are almost exclusively evaluated on datasets with none or hardly any attention switches, feature vectors that are recorded in close succession are extremely likely to belong to the same class. Although the feature drift itself is completely uninformative, it thus artificially clusters features that belong to the same attention class if attention to the same speaker is sustained for a long time \cite{rotaruWhatAreWe2024}. It then suffices to know (a good prediction of) the label of a single feature per cluster to achieve a strong performance, even if the predictions themselves are barely modulated by the actual attention process. 

    The most blatant example of this process in practice is the usage of random cross-validation to train direct-classification models that directly assign an attention label to an EEG snippet (e.g. attend left versus attend right) without correlating the EEG explicitly with the speech stimuli. Random cross-validation assigns each EEG window randomly to either the train or test set. This gives the classifier access to many ground-truth labels of features recorded closely before and after each test window. If there is a significant amount of temporal drift present (which is the case for EEG-based classification, as we will discuss later), the classifier can then simply abuse the temporal proximity of features to 'copy' the training label to the test label \cite{rotaruWhatAreWe2024,yanOverestimatedPerformanceAuditory2025}. 

    Another example is the usage of unsupervised, adaptive algorithms. These algorithms iteratively re-train the AAD classifier using generated pseudo-labels \cite{Geirnaert2021Unsup, Geirnaert2022Unsup, heintzUnsupervisedEEGbasedDecoding2025}. If the features are conditionally dependent, the AAD classifier gets biased to copy the most recent pseudo-labels in the new prediction. Although this is not problematic as such (the pseudo-labels are generated in an unsupervised way), it does create an indirect form of smoothing, which may decrease the temporal resolution as discussed above.

    \subsection{A decreased postprocessing potential}
    Raw per-window AAD predictions are rarely directly used in gain control algorithms. Instead, they are generally first fused over time using some form of postprocessing \cite{Presacco2019, hjortkjaerRealtimeControlHearing2025, Geirnaert2020a,heintzProbabilisticGainControl2024, heintzPostprocessingEEGbasedAuditory2025}. This significantly decreases the misclassification rate, at the cost of a slightly decreased temporal resolution. However, when consecutive AAD predictions are already conditionally dependent on each other, the potential benefit of such postprocessing algorithms is significantly reduced \cite{heintzPostprocessingEEGbasedAuditory2025}.

    For example, consider a naive postprocessing where the decision scores of an AAD algorithm are smoothed using a weighted moving average filter. When these predictions are independent, averaging reduces the variance and thus improves the accuracy. However, when there are significant conditional dependencies, this regression to the mean happens much slower. Further improving the accuracy comes thus at a much greater cost in temporal resolution. Therefore, even if an AAD algorithm with independent predictions has a much lower accuracy than its counterpart with dependent predictions, it may still outperform the dependent predictions after postprocessing \cite{heintzPostprocessingEEGbasedAuditory2025}. 

    \section{Using information to merge performance and independence}
    \label{sec:informationTheory}
    To avoid all the potential pitfalls of evaluating conditionally dependent predictions on steady state data, we would ideally wish to evaluate each AAD algorithm on a dataset where past data cannot be exploited to predict the current state. In other words, the measured participant should constantly switch attention, choosing a new target speaker at random for every window. However, this is practically infeasible: it would quickly lead to extreme exhaustion, where even the most motivated participant would drop off in mere minutes. Alternatively, it is possible to solely focus on the AAD accuracy achieved on predictions generated just after an attention switch to estimate the temporal resolution, but this would mean that only a very small percentage of a recording can be used for evaluation, again leading to noisy performance metrics. Furthermore, most existing AAD datasets don't have any attention switches, except for a few where switches in attention remain very sparse \cite{rotaruAudiovisualGazecontrolledAuditory2024}. Finally, even if datasets with an abundance of attention switches would exist in the future, the exact moment a participant switches attention may always be off by a few seconds relative to the switching cue that defines the ground truth. As AAD algorithms improve, this label uncertainty would become an important source of variance, making exact comparisons hard.

    However, it is possible to achieve a similar goal by analysing the performances through the lens of information theory. In general, we can model the AAD problem as a process with a hidden attention state $y(n)$ (the identity of the attended speaker at window $n$) and scores $\Vec{f}(n)$. While the scores can be anything, ranging from raw samples of a recorded EEG to classifier outputs (or any feature representation in between), we will (without loss of generality) specifically focus on the classification scores generated by AAD algorithms. In this case, the scores $\Vec{f}(n)$ can be viewed as a probability vector for each speaker being the attended speaker, or a one-hot vector with a hard decision. Given the hidden state $y(n)$, scores $\Vec{f}(n)$, and past scores $\Vec{f}_{past}(n)=[\Vec{f}^\top(n-\tau)\dots\Vec{f}^\top(n-1)]^\top, \ \tau > 0$, we will study four key metrics: Mutual Information (MI), Conditional Self-Dependence (CSD), Incremental MI (IMI), and relative IMI (rIMI), which will be explained in the remaining subsections, after explaining some generic information-theroetic concepts in the next subsection.

    \subsection{Entropy and information}
    \label{sec:entropy}
    \textbf{Entropy} is a measure of uncertainty, and is measured in bits. The entropy of a variable $x$ is written as $H(x)$. If $H(x)=\SI{0}{\bit}$, the value of $x$ is known with absolute certainty. If $H(x)=\SI{1}{\bit}$, there is as much uncertainty about the value of $x$ as about the outcome of a coin toss. Formally, entropy is defined as: 
    
    \begin{equation}
        H(x) = -\int \p(x)\log_2(\p(x)) \mathrm{d} x,
    \end{equation}
    where $\p(x)$ is the probability density function of $x$.
    
    \textbf{Conditional entropy} $H(x|y)$ represents the amount of uncertainty about the value of $x$ if $y$ is known. If $x$ and $y$ are independent, $H(x|y)=H(x)$: knowledge about the value of $y$ does not reduce the uncertainty about the value of $x$. When there is a dependence between $x$ and $y$, the conditional entropy can be computed as:
    \begin{equation}
        H(x|y) = -\int \p(x,y)\log_2\left(\frac{\p(x,y)}{\p(x)}\right) \mathrm{d} x \mathrm{d}y.
    \end{equation}
    
    \textbf{Mutual Information (MI)} $I(x;y)$ represents how knowledge of $y$ reduces the uncertainty about the value of $x$, and vice versa. Formally, the mutual information is defined as:
    \begin{align}
        I(x;y) &= H(x) - H(x|y)\\
        &= H(y) - H(y|x).
        \label{eq:MI-MutualInformation}
    \end{align}
    
    \textbf{Conditional Mutual Information (CMI)} $I(x;y|z)$ represents how knowledge of $y$ reduces the uncertainty about the value of $x$ (or vice versa) if $z$ is already known. If $x$ or $y$ is strongly correlated with $z$, the CMI will be low. If $x$ and $y$ are completely independent of $z$, the CMI is identical to the MI $I(x;y)$. The conditional mutual information is defined as:
    \begin{equation}
        I(x;y|z) = I(x;y,z) - I(x;z),
        \label{eq:MI-CMI}
    \end{equation}
    where $I(x;y,z)$ represents the mutual information between $x$ and the stacked variable $(y,z)$. 

    \subsection{Mutual Information (MI)}
    In the context of AAD, we  are typically interested in the quantity $I(\Vec{f}(n);y(n))$, which is the MI between the attention state $y(n)$ and the observation $\Vec{f}(n)$. It measures how well you can predict the attention state given the observations and vice versa. This metric is very similar to the accuracy of an AAD algorithm and is regularly used to compute the so-called Information Transfer Rate (ITR) \cite{Geirnaert2020}. Crucially, $I(\Vec{f}(n);y(n))$ does \textbf{not} take into account whether consecutive observations are dependent, and thus suffers from the same problems as accuracy.
    
    \subsection{Conditional Self-Dependence (CSD)} 
    We define the conditional self-dependence (CSD) as $I(\Vec{f}(n);\Vec{f}_{past}(n)|y(n))$, which is the CMI between current observation $\Vec{f}(n)$ and previous observations $\Vec{f}_{past}(n)$ given the underlying attention state $y(n)$. The new observation is conditionally independent from previous observations if $I(\Vec{f}(n);\Vec{f}_{past}(n)|y(n))=0$, i.e., if the previous observations $\Vec{f}_{past}(n)$ are unrelated to the current observation $\Vec{f}(n)$, other than their shared attention state $y(n)$. This metric is similar to a partial correlation but also accounts for nonlinear dependencies \cite{inceStatisticalFrameworkNeuroimaging2017}.
    
    \subsection{Incremental Mutual Information (IMI)}
    We define the incremental mutual information (IMI) as $I(\Vec{f}(n);y(n)|\Vec{f}_{past}(n))$, which is the CMI between current observation $\Vec{f}(n)$ and the attention state $y(n)$ given previous observations $\Vec{f}_{past}(n)$. This metric is higher when the new observation provides more previously unavailable information about the identity of an attended speaker. This metric therefore allows us to separate novel information on the attention state from information that is carried over from previous predictions. If there is no CSD between previous and current predictions, the IMI is exactly equal to the MI.
    
    \subsection{relative Incremental Mutual Information (rIMI)}
    We define the rIMI in \eqref{eq:rIMI} as the fraction of the remaining uncertainty about the current attention state $y(n)$ that is removed by the current observation $\Vec{f}(n)$.
    
    \begin{equation}
        rIMI \triangleq \frac{I(\Vec{f}(n);y(n)|\Vec{f}_{past}(n))}{H(y(n)|\Vec{f}_{past}(n))}.
        \label{eq:rIMI}
    \end{equation}

     A perfect model has an rIMI of $1$, while a random model or a model that merely repeats previous predictions has an rIMI of $0$. This makes the rIMI much easier to interpret than the IMI, which has a variable upper bound dictated by the total amount of remaining state uncertainty $H(y(n)|\Vec{f}_{past}(n))$. 
     
	\section{Estimating information-theoretic metrics}
	\label{sec:MI-MutualInformation}
    In this section, we slowly build towards the practical computation strategies for the four key metrics (MI, CSD, IMI, rIMI) that we will use to evaluate AAD algorithms, being the core focus of this work. To differentiate between general theoretical concepts and implementations that are specific to AAD, we use $\Vec{x}(n)$ for observations in general, and $\Vec{f}(n)$ specifically for AAD classification scores for each window $n$. This allows us to clearly separate theoretical concepts where we make Gaussianity assumptions on the distribution of $\Vec{x}(n)$ from the practical implementations without this assumption at the end of this section. 
    
	\subsection{Gaussian entropy}
    \label{sec:MI-GaussianEntropy}
	A key downside of information-theoretic metrics is that they are often hard to estimate. Estimating the entropy $H(\Vec{x}(n))$ of a variable $\Vec{x}(n)$ requires knowledge of the distribution $\p(\Vec{x}(n))$, which is generally unknown. While these distributions can be estimated using binning methods, such approximations are often too inaccurate to provide a reliable estimation of the entropy \cite{inceStatisticalFrameworkNeuroimaging2017}. However, in some rare cases, the entropy $H(\Vec{x}(n))$ has a closed-form solution. One important example is when $\Vec{x}(n)$ is sampled from a $K$-dimensional multivariate Gaussian distribution, i.e. $\Vec{x}(n) \sim \mathcal{N}(\bm{\mu},\Sigma_\Vec{x})$ \cite{inceStatisticalFrameworkNeuroimaging2017}:
	
	\begin{equation}
		H(\Vec{x}(n)) = \frac{1}{2} \log_2\left((2\pi e)^K | \Sigma_\Vec{x} |\right).
	\end{equation}
	The entropy of a Gaussian variable thus only depends on its dimensionality $K$ and the determinant $| \Sigma_\Vec{x} |$ of its covariance matrix $\Sigma_{\Vec{x}}$.
	
	This result can easily be expanded to compute the conditional entropy $H(\Vec{x}(n)|y(n))$ of a variable $\Vec{x}(n)$ which is sampled from a Gaussian mixture (instead of a single Gaussian), where the discrete state variable $y(n)$ indicates from which Gaussian distribution $\Vec{x}(n)$ is sampled, i.e. $\Vec{x}(n)|(y(n) = i) \sim \mathcal{N}(\bm{\mu}_i,\Sigma_{\Vec{x},i})$:
	\begin{align}
		H(\Vec{x}(n)|y(n)) 	&= \sum_i p_i H(\Vec{x}(n)|y(n)=i)\\
		&= \sum_i p_i \frac{1}{2} \log_2\left((2\pi e)^K | \Sigma_{\Vec{x},i} |\right),
        \label{eq:MI-conditionalEntropy}
	\end{align}
	with $p_i$ the probability that $y(n)=i$.
	
	While there is no closed-form solution for the unconditioned entropy $H(\Vec{x}(n))$ of a variable $\Vec{x}(n)$ sampled from a Gaussian mixture, several fast algorithms exist that can accurately approximate $H(\Vec{x}(n))$ for Gaussian mixtures. In this paper, we specifically use the algorithm proposed by Huber et al. \cite{huberEntropyApproximationGaussian2008}.
	
	These identities allow us to efficiently estimate the four metrics used in this paper. Using \eqref{eq:MI-MutualInformation}, we can compute the \textbf{mutual information} (MI) between a mixed Gaussian variable $\Vec{x}(n)$ and the underlying state $y(n)$:
	\begin{equation}
		I(\Vec{x}(n);y(n)) = H(\Vec{x}(n)) - \sum_i p_i \frac{1}{2} \log_2\left((2\pi e)^K | \Sigma_{\Vec{x}_i} |\right).
        \label{eq:MI-MIestimation}
	\end{equation}
	
	This estimate for the mutual information can then be used to estimate the \textbf{incremental mutual information} $I(\Vec{x}(n);y(n)|\Vec{x}_{past}(n))$ based on \eqref{eq:MI-CMI}:
	\begin{align}
	    &I(\Vec{x}(n);y(n)|\Vec{x}_{past}(n)) = \nonumber\\
        &\quad I(\Vec{x}(n),\Vec{x}_{past}(n);y(n)) - I(\Vec{x}_{past}(n);y(n)).
		\label{eq:MI-CMIestimation}
	\end{align}
    The first term is the mutual information between the collection of both past and present observations $[\Vec{x}^\top(n),\Vec{x}^\top_{past}(n)]^\top$ and the attention state $y(n)$. The second term models how much we already knew about $y(n)$ from past observations. 

    Using the definition \eqref{eq:rIMI}, the \textbf{relative incremental mutual information} is defined as the fraction between the IMI and the remaining uncertainty $H(y(n)|\Vec{x}_{past}(n))$ about the state $y(n)$ given previous observations $\Vec{x}_{past}(n)$. The numerator in \eqref{eq:rIMI} is the IMI, which is computed as specified in \eqref{eq:MI-CMIestimation}. The denominator of \eqref{eq:rIMI} becomes $H(y(n)|\Vec{x}_{past}(n))$, which can be computed from \eqref{eq:MI-MutualInformation} as:
    
    \begin{align}
        & H(y(n)|\Vec{x}_{past}(n)) = H(y(n)) - I(\Vec{x}_{past}(n);y(n))\\
        & H(y(n)) \triangleq -\sum_i p_i \log_2 \left( p_i \right),
    \end{align}
    with $p_i$ the prior probability that $y(n) = i$.
    
	Finally, the \textbf{conditional self-dependence} $I(\Vec{x}(n);\Vec{x}_{past}(n)|y(n))$ between observations $\Vec{x}(n)$ and $\Vec{x}_{past}(n)$ given state $y(n)$ can be computed using \eqref{eq:MI-conditionalEntropy}:
	\begin{align}
		&I(\Vec{x}(n);\Vec{x}_{past}(n)|y(n)) 	\\
        &\quad = H(\Vec{x}(n)|y(n)) + H(\Vec{x}_{past}(n)|y(n))\nonumber\\
        &\qquad- H(\Vec{x}(n),\Vec{x}_{past}(n)|y(n))\\
		&\quad = \sum_i p_i \frac{1}{2} \log_2\left(\frac{|\Sigma_{\Vec{x},i}|| \Sigma_{\Vec{x}_{past},i} |}{|\Sigma_{\Vec{x}\Vec{x}_{past},i}|}\right).
        \label{eq:MI-conditionalDependenceEstimation}
	\end{align}
	Here, $H(\Vec{x}(n),\Vec{x}_{past}(n)|y(n))$ and $\Sigma_{\Vec{x}\Vec{x}_{past},i}$ represent the conditional entropy and the covariance matrix of the joint variable $[\Vec{x}(n)^\top \Vec{x}^\top_{past}(n)]^\top$, respectively. 
    
	\subsection{The copula trick}
	Naturally, the scores $\Vec{f}(n)$ generated by an AAD algorithm are not necessarily sampled from a mixed Gaussian distribution. However, it is possible to transform a non-Gaussian variable $\Vec{f}(n)$ to a Gaussian variable $\Vec{f}_\mathcal{G}(n)$ using the Gaussian copula \cite{inceStatisticalFrameworkNeuroimaging2017}. This transformation works as follows:
	
	\begin{itemize}
		\item For each discrete label $i$ and each dimension $k$ of $\Vec{f}$, rank all values $f_k(n)$ where $y(n)=i$ across all samples $n=[0,N]$.
		\item Replace each value $f_k(n)$ by its ranking.
		\item Normalise the rankings per dimension, such that they are constrained between $[\frac{1}{R+1},\frac{R}{R+1}]$, with $R$ the maximal ranking.
		\item $f_{\mathcal{G},k}(n)$ is then the inverse standard normal cumulative density function of the normalised rank of $f_k(n)$.
	\end{itemize}
	
	Ince et al. showed that both the mutual information $I(\Vec{f}_\mathcal{G}(n);y(n))$ and the conditional self-dependence $I(\Vec{f}_\mathcal{G}(n);\Vec{f}_{\mathcal{G},past}(n)|y(n))$ of the Gaussian transformation provide lower bounds for the mutual information and conditional self-dependence of $\Vec{f}(n)$ \cite{inceStatisticalFrameworkNeuroimaging2017}. This allows us to use these fast Gaussian Copula Mutual Information estimators to estimate the key information-theoretic metrics defined in Section \ref{sec:MI-GaussianEntropy}. 
    
	\section{Experiments}
	\label{sec:MI-Experiments}
    To demonstrate how ignoring conditional self-dependence (e.g., by solely using the accuracy) can lead to misleading results, and how to correct for this using the (r)IMI metric, we outline the details of several experiments in this section. The results of these experiments are discussed in Section \ref{sec:MI-Results}.
    
	\subsection{Dataset}
	All AAD algorithms were benchmarked on the audio-visual gaze controlled (AV-GC) dataset presented by Rotaru et al. \cite{rotaruWhatAreWe2024,rotaruAudiovisualGazecontrolledAuditory2024}. This AAD dataset was developed with the purpose of removing potential shortcuts due to eye artifacts in the EEG that are correlated with the locus of auditory attention \cite{rotaruWhatAreWe2024}, which is a shortcut that is almost always exploited by spatial or direct-classification AAD algorithms that directly classify the locus of auditory attention (i.e., left or right speaker attended). In this dataset, 13 participants are instructed to listen to one of two competing speakers. The competing speakers were located on the left and right sides of the head. In the meantime, the eye gaze was controlled using various methods, depending on the trial condition: 
    \begin{itemize}
		\item \textbf{Moving video (MV)}: the participants were instructed to look at a randomly moving video of the attended speaker.
		\item \textbf{Moving target noise (MTN)}: the participants were instructed to look at a randomly moving cross-hair.
        \item \textbf{Static Video (SV)}: the participants were instructed to look at a video of the attended speaker located at the same side as the position of the attended speaker. This introduces a majore gaze-related artifact towards the side of auditory attention that can be used as a shortcut to perform AAD.
		\item \textbf{No visuals (NV)}: the participants were instructed to fixate on an imaginary point in front of them. For some participants, this still results in a slight yet exploitable gaze bias towards the side of auditory attention.
	\end{itemize}
	Although all conditions are used for training, we only use the MTN, NV and MV conditions for validation to ensure that gaze-related artifacts or biases in the EEG data cannot be used to decode the auditory attention. 
	
	Each trial lasts \SI{10}{\min}, with an attention switch after \SI{5}{\min} (either left-to-right or right-to-left). Each condition is repeated twice: once with the left speaker first attended and once with the right speaker first attended. For most participants, there are thus a total of eight \SI{10}{\min} trials. However, the MTN trials are not present in the first two participants. 
	
	The EEG is recorded with a 64-channel EEG BioSemi ActiveTwo system. The EEG and audio are then preprocessed using the preprocessing framework proposed by Biesmans et al. \cite{Biesmans2017}. For more information about the dataset, we refer to \cite{rotaruWhatAreWe2024}.
	
	\subsection{AAD algorithms}
	To investigate how conditional self-dependence influences the performance of various AAD algorithms, we evaluated several AAD algorithms from the stimulus reconstruction and the direct-classification family. It is noted that the goal of this work is not to provide a benchmark across all existing AAD algorithms, but merely to illustrate how existing performance metrics are flawed when used in solitude, which is why we only selected a representative subset of AAD algorithms from both families.
    
   Concerning the stimulus reconstruction algorithms, we evaluated the Least-Squares (LS) algorithm proposed by O'Sullivan et al. \cite{OSullivan2014}, the Canonical Correlation Analysis (CCA) algorithm with linear discriminant analysis (LDA) classifier proposed by de Cheveigné et al. \cite{DeCheveigne2018}, and the non-linear stimulus decoding algorithm AADNet proposed by Nguyen et al. \cite{nguyenAADNetEndtoEndDeep2025}. 
   
   Concerning the spatial/direct-classification AAD algorithms, we evaluated the linear common spatial patterns (CSP) algorithm with LDA classifier proposed by Geirnaert et al. \cite{Geirnaert2020CSP}, an adaptive (oracle) version of the latter, and the nonlinear Dual Attention Refinement Network (Darnet) proposed by Yan et al. \cite{yanDARNetDualAttention2024}. The adaptive CSP algorithm uses oracle labels to update the LDA classifier over time, and should therefore be viewed as an upper-bound for the performance of an unsupervised adaptive CSP version\footnote{This could be achieved by using pseudo-labels generated by the unsupervised algorithms in \cite{Geirnaert2021Unsup} or \cite{Heintz2023Unbiased}.}. Such an adaptive version of the algorithm would be able to partially compensate for feature drift in the CSP feature space, which is known to severely deteriorate performance \cite{rotaruWhatAreWe2024}. The exact algorithm used to update the LDA classifier is specified in Appendix \ref{app:adaptive CSP algorithm}.

   Each algorithm was trained and tested using leave-one-trial-out (LOTO) cross-validation (CV). We then measure the accuracy, mutual information (MI), relative incremental mutual information (rIMI) and conditional self-dependence (CSD) per trial. 
   
   All information-theoretic metrics were derived by first combining the algorithm's classification scores $\Vec{f}(n)$ across all windows and trials per subject, and then estimating the appropriate distributions for each subject using the Gaussian Copula, as explained in Section \ref{sec:MI-MutualInformation}. All windows $n$ where there was at least one attention switch between $n-\tau$ and $n$ were removed. This ensures that the past data contained in $\Vec{f}_{past}(n)$ are always relevant to predict $y(n)$, and avoids discrepancies when one would compare the (r)IMI across datasets with different amounts of attention switches.
   
   The relevant hyperparameters for LS, CCA and CSP are shown in Table \ref{tab:MI-AADHyperParam}. For all other algorithms, we used the exact hyperparameters proposed by the authors.
	
	\begin{table}[h]
		\centering
		\begin{tabular}{l l c }
			\textbf{Algorithm} & \textbf{Hyperparameter} & \textbf{Value}\vspace*{0.3cm}\\
            All & Window length & \SI{1}{\second}\\
			LS 	& EEG lags $L$ 				& \SIrange{0}{250}{\milli\second}\\
			\hline
			& EEG lags $L_{eeg}$ 		& \SIrange{0}{250}{\milli\second}\\
			CCA & Envelope lags $L_{stim}$	& \SIrange{-250}{0}{\milli\second}\\
			& Number of transformations $K$ & 5\\
			\hline 
			CSP & Frequency bands			& Full band (\SIrange{0}{32}{\hertz})\\
			& Number of transformations $K$ & 5		
		\end{tabular}
		\caption{The hyperparameters used in each AAD algorithm.}
		\label{tab:MI-AADHyperParam}
	\end{table}

    For LS, we used the difference between the attended and unattended correlation as $\Vec{f}(n)$. For all other models, we used the one-dimensional classification score obtained at the output of the model to construct $\Vec{f}(n)$. 
	
	\subsection{Estimation of the information metrics}
	The mutual information and the conditional self-dependence were directly estimated using the GCMI toolbox \cite{inceStatisticalFrameworkNeuroimaging2017}. The IMI and rIMI are then estimated using the identities specified in Section \ref{sec:MI-MutualInformation}. These identities are also estimated with the GCMI toolbox. 
	
	The previous AAD predictions $\Vec{f}_{past}(n)$ used to compute the (r)IMI and conditional self-dependence, consist of a vector that stacks all scores from all decision windows in the last \SI{5}{\second} (excluding the current prediction) for window lengths $\leq$\SI{5}{\second}. For longer window lengths, only the last AAD prediction was used. This allows us to correct for both short-term and long-term dependencies between scores. Ideally, as many previous predictions as possible should be used to correct for dependencies that persist for longer. However, this would significantly increase the dimensionality of $\Vec{f}_{past}(n)$, where the curse of dimensionality would negatively affect the quality of the estimated information-theoretic metrics.
	
	\subsection{Experiments}
	In total, we will study the AAD algorithms through an information-theoretic lens based on four key experiments. However, to facilitate the analysis, and reduce clutter, only the first experiment is performed on all AAD algorithms. The other experiments are performed only with the LS and CSP algorithms. The observed trends for LS and CSP are representative for the other AAD algorithms within their respective classes (i.e., stimulus reconstruction and direct-classification AAD). 

    \paragraph{The relationship between accuracy and rIMI on synthetic data}
    To facilitate the interpretation of the novel rIMI metric, we first investigate the relationship between the accuracy and the rIMI on synthetic data. We consider two datasets. In the first dataset, 1000 observations $f(n)$ are independently sampled from one of two one-dimensional Gaussian distributions. The first 500 observations are sampled from the first distribution, the second 500 observations from the second. Both distributions have variance $\sigma_{1,2}=1$. The distance between the class averages $\|\mu_1-\mu_2\|$ is iteratively increased from 0 to 4 to artificially improve the decoding accuracy. The first half of the observations receive label $y(n)=1$, the second half $y(n)=2$.

    The second dataset is directly constructed from the first, with observations $f'(n) = f(n) + 1/3 \sum_{t=0}^2 f(n-t)$. The current observation thus also depends on the two previous observations. 
    
    At each iteration of the dependent and independent datasets, the observations were classified to the class $i$ that minimises $|f(n) - \mu_i|$. We then compute the accuracy and rIMI. The rIMI is conditioned on the past 3 samples. We then compare how the accuracy and rIMI changes for both the first dataset with independent samples, and the second dataset with conditionally dependent samples.
    
	\paragraph{An information-theoretic comparison of AAD algorithms}
	The accuracy and the four information-theoretic metrics are computed on the AV-GC dataset for each AAD algorithm, window length and participant separately. For spatial AAD algorithms, both the leave-one-trial-out cross-validation and random cross-validation results are reported.
    
	\paragraph{The accuracy and AAD scores around a switch}
	To better understand how spatial and stimulus reconstruction AAD algorithms behave around an attention switch, we compute the average evolution of the accuracy and classification scores $\Vec{f}(n)$ over time around a switch of attention within the data. Ideally, the scores should change immediately after an attention switch and be significantly different before and after the switch. The accuracy should remain stable across the entire trial, with potentially only a short dip in accuracy around the switch, as there is always some time between an instructed switch and the participant actually switching attention. To limit noise, a relatively long \SI{5}{\second} window is used to compute the classification scores. However, using shorter window lengths did not affect the main conclusions.

    The reported accuracies are the average accuracies across participants and trials at one specific moment in time. The scores are first aligned so that the cued attention switch always occurs at the same time, and the left speaker is first attended. When the opposite is true, the scores are multiplied by -1 to ensure alignment. The scores are then averaged over all trials and subjects. Finally, this result is normalised to facilitate the comparison between algorithms. 

	\paragraph{The origin of dependencies}
	In the last experiment, we study the origin of the prediction dependencies. As will be demonstrated in Section \ref{sec:MI-Results}, the conditional self-dependency between consecutive predictions is significantly higher for spatial direct-classification AAD algorithms than stimulus reconstruction algorithms. We hypothesise that this is because covariance structure used by stimulus reconstruction algorithms is less dependent than the covariance structure used by direct-classification algorithms. 
    
    Stimulus reconstruction algorithms generate predictions based on the interactions between the EEG signal $\Vec{m}(t)$ and the speaker envelopes $\Vec{s}(t)$, within a certain window $n$. For instance, the LS decoder $\Vec{d}_{LS}$ generates features proportional to $\Vec{d}_{LS}^\top\Vec{m}(n)s(n)^\top \propto \Vec{d}^\top R_{ms}(n)$, with $R_{ms}(n)$ the sample crosscorrelation matrix in a window $n$. Similarly, CSP decoders $\Vec{d}_{CSP}$ produce features proportional to $\Vec{d}_{CSP}^\top R_{mm}(n) \Vec{d}_{CSP}$, with $R_{mm}(n)$ the sample autocorrelation matrix in window $n$.
    
    We therefore study the conditional dependencies $I(R_{mm}(i,j,n);R_{mm}(i,j,n-1)|y(n))$ and $I(R_{ms}(i,j,n);R_{ms}(i,j,n-1)|y(n))$ for each $i$ and $j$. If this dependency is high, this means that consecutive predictions are (partially) made based on data that haven't significantly changed. Assuming the prediction model is linear or sufficiently smooth, these dependent data samples will then be translated into conditionally dependent predictions.
	
	\section{Results and discussion}
	\label{sec:MI-Results}
    \subsection{The relationship between accuracy and rIMI on synthetic data}
\begin{figure}[ht]
		\centering
		\includegraphics[width=0.9\columnwidth]{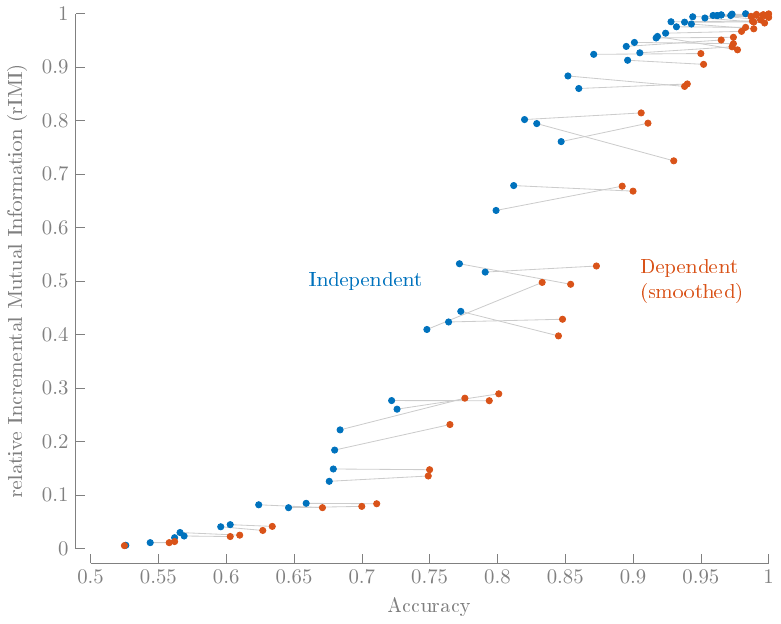}
		\caption[The relation between the accuracy and rIMI.]{The relation between the accuracy and rIMI for increasing clas separation in both the independent and dependent dataset. While temporal smoothing can improve the accuracy, it has on average no effect on the rIMI.}
		\label{fig:MI-rIMIVSAcc}
	\end{figure}

    Figure \ref{fig:MI-rIMIVSAcc} shows the rIMI versus accuracy, each showing an increasing trend when the class separation is increased. The additional smoothing that happens in the dependent dataset strongly affects the accuracy, but not the rIMI. This is desired: the rIMI only measures how much \textit{new} information an observation contains. Temporal smoothing only transfers information that was already available to the current observation, and thus cannot increase the amount of new information that an observation contains, i.e., the rIMI should not change between the two datasets. 
    
	\subsection{An information-theoretic comparison of AAD algorithms}
	\label{sec:MI-ITcomparison}
\begin{figure*}[t]
    \centering
    \begin{subfigure}[t]{0.48\textwidth}
        \centering
        \includegraphics[width=\linewidth]{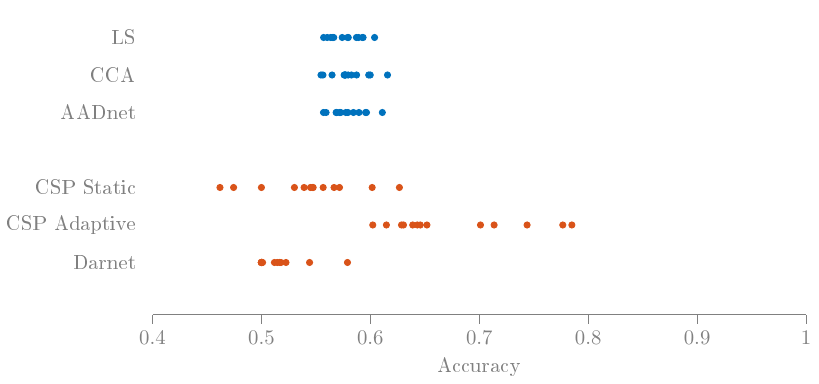}
        \caption{Accuracy (\textbf{Can be misleading)}}
    \end{subfigure}
    \hfill
    \begin{subfigure}[t]{0.48\textwidth}
        \centering
        \includegraphics[width=\linewidth]{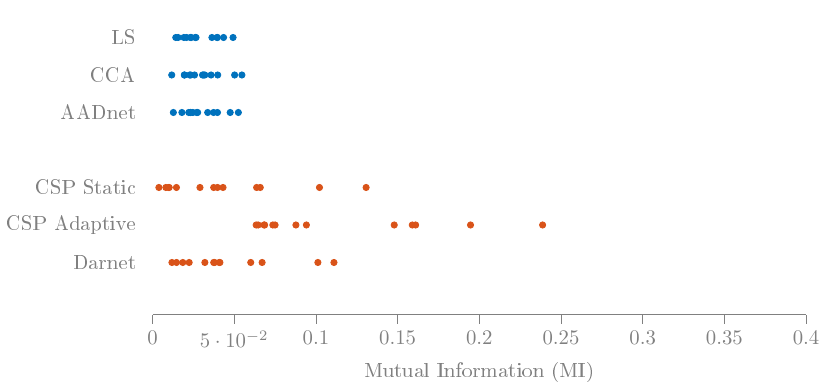}
        \caption{Mutual Information (MI) \textbf{(Can be misleading)}}
    \end{subfigure}

    \vspace{0.5em}

    \begin{subfigure}[t]{0.48\textwidth}
        \centering
        \includegraphics[width=\linewidth]{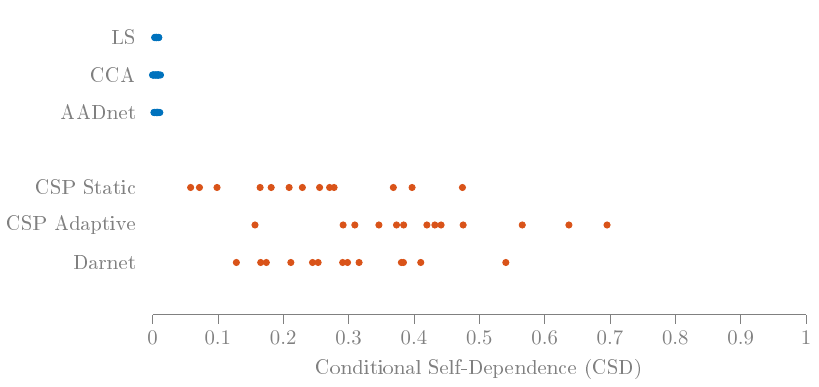}
        \caption{Conditional Self-Dependence (CSD)}
    \end{subfigure}
    \hfill
     \begin{subfigure}[t]{0.48\textwidth}
        \centering
        \includegraphics[width=\linewidth]{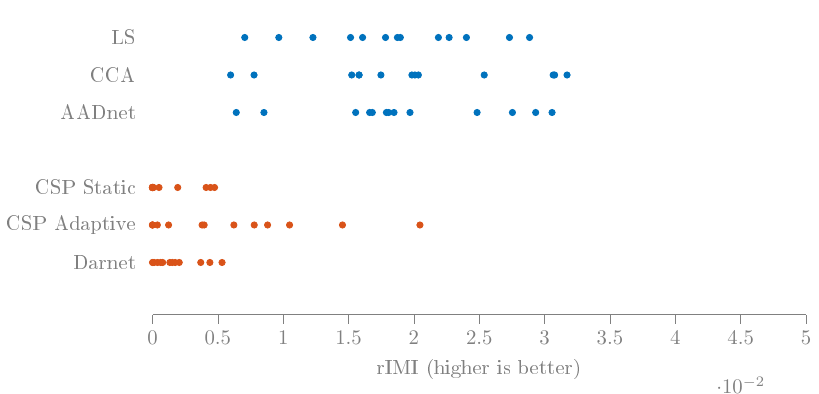}
        \caption{relative Incremental Mutual Information (rIMI)}
    \end{subfigure}

    \caption{The Accuracy, Mutual Information (MI), conditional self-dependence (CSD) and relative Incremental Mutual Information (rIMI) for various stimulus reconstruction and spatial AAD algorithms on \SI{2}{\second} windows. While CSP with adaptive classifier appears to perform better at first glance, based on traditional metrics like accuracy and MI (used in ITR), its performance is severely overestimated due to the conditional self-dependence between consecutive predictions. The rIMI corrects for this dependence by only measuring how much new information a model generates per window. This is significantly higher for all stimulus reconstruction algorithms.}
    \label{fig:ITmetrics}
\end{figure*}

    Figure \ref{fig:ITmetrics} shows the accuracy, Mutual Information (MI), Conditional Self-Dependency (CSD) and relative Incremental Mutual Information (rIMI) for each tested model on \SI{2}{\second} windows. The accuracy and MI (as used in the popular information transfer rate metric) do not take into account that predictions may be dependent.
    
    The CSD measures to what extent consecutive predictions are dependent. A higher dependence means that a larger portion of the prediction is repeated from earlier predictions. Here we see a clear difference between stimulus reconstruction methods (LS \cite{OSullivan2014}, CCA \cite{DeCheveigne2018} and AADnet \cite{nguyenAADNetEndtoEndDeep2025}), and direct-classification AAD methods such as CSP with static and adaptive classifiers \cite{Geirnaert2020CSP}, and Darnet \cite{yanDARNetDualAttention2024}. The CSD is  near-zero for the stimulus reconstruction algorithms, and much higher for the direct-classification methods. While a high CSD is not necessarily problematic, it does imply that (1) the accuracy cannot be easily improved through postprocessing, such as temporal smoothing \cite{heintzPostprocessingEEGbasedAuditory2025}, (2) the effective temporal resolution may be lower than what the window length suggests, and (3) that the model may be prone to overfitting on temporal fingerprints. This means that the model simply decodes \textit{when} a certain test window was recorded, and infers from nearby training data what its corresponding class is. This is viable strategy when attention switches are rare and there is training data available that was recorded shortly before or after the test window (e.g., when using random cross-validation, adaptive decoders, or when some training trials were recorded shortly before or after the test trial).
    
    The rIMI specifically measures the fraction of remaining uncertainty about the identity of the attended speaker that is removed by a new prediction. A rIMI of 0.02 means that each prediction approximately removes 2\% of the remaining uncertainty about the attention state. A model that merely parrots previous predictions has a rIMI equal to 0, no matter its accuracy, as it fails to generate any new information. A good model should obtain both a high accuracy and a high rIMI. When the accuracy is high, but the rIMI is low (such as CSP with adaptive LDA), this indicates that the high accuracy is primarily driven by repeating previous predictions, rather than generating actual new information. This is problematic from the moment there is an actual switch in attention, as we will see later on. We further explore how a model such as adaptive CSP can obtain such a high accuracy, but low rIMI, and why this is a clear sign of overfitting in Section \ref{subse:MI-accuracyInTrial}.
    
\begin{figure*}[t]
    \centering
    \begin{subfigure}[t]{0.48\textwidth}
        \centering
        \includegraphics[width=\linewidth]{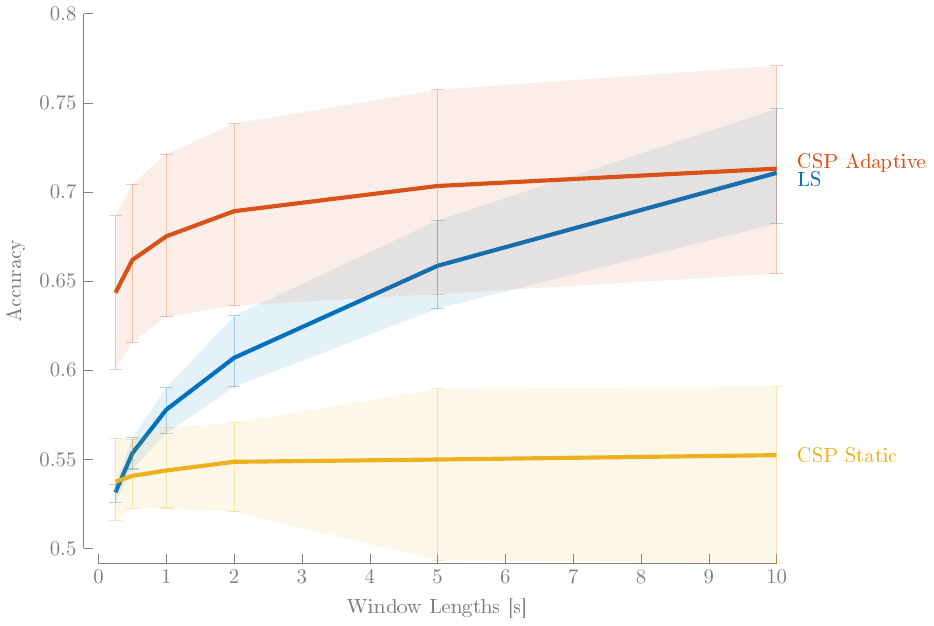}
        \caption{Accuracy (\textbf{Can be misleading})}
        \label{fig:accWL}
    \end{subfigure}
    \hfill
    \begin{subfigure}[t]{0.48\textwidth}
        \centering
        \includegraphics[width=\linewidth]{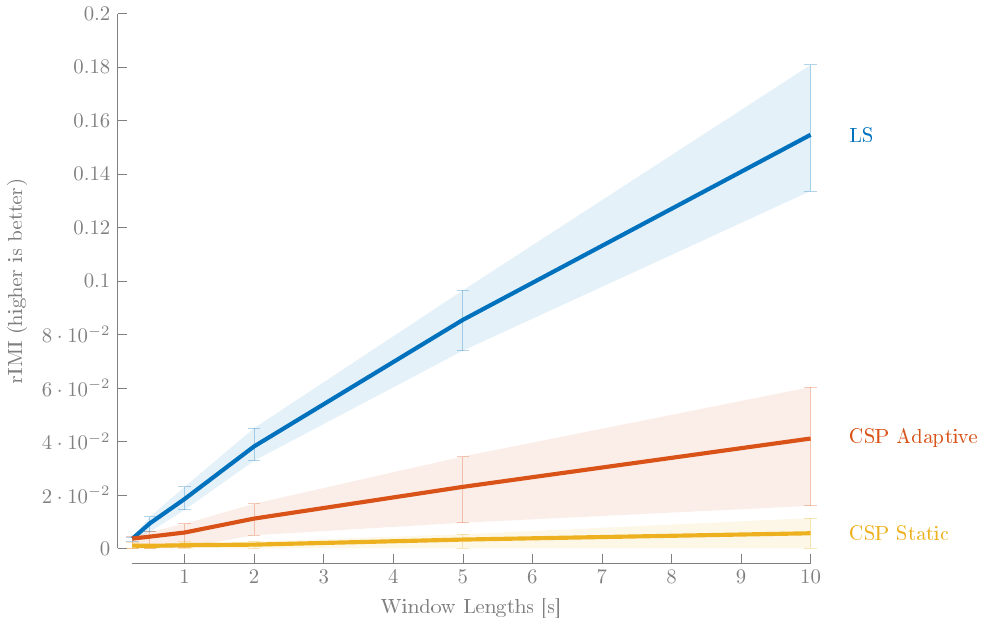}
        \caption{relative Incremental Mutual Information (rIMI)}
    \end{subfigure}
    \caption{The accuracy and rIMI per window length. Although the spatial CSP model with adaptive classifier seems to outperform the LS stimulus reconstruction model, it generates significantly less new information per window than LS.} 
    \label{fig:AccRimiWL}
\end{figure*}

    In Figure \ref{fig:AccRimiWL}, we also report the accuracy and rIMI in function of the window length for LS and CSP with adaptive and static classifier. The figure demonstrates that there is a limit to reducing the window length in a meaningful way. Although the accuracy stays significantly above chance level for all models on very short decision windows, the rIMI quickly approaches 0. On such short windows, the biological processes that are captured vary too slowly to truly capture new information in consecutive windows. While spatial AAD algorithms are often reported to achieve high accuracies on \SI{0.1}{\second} windows \cite{Zhang, suSTAnetSpatiotemporalAttention2022, yanDARNetDualAttention2024}, this does not mean that these algorithms would actually be able to detect several sub-second attention switches. At such time scales, information is already smeared over several windows by various preprocessing filters, even if we ignore the inherent latencies in the biological processes. So while the accuracy and MI may remain high (when the classifier is trained on trial-specific data), the actual rate of new information per window nears 0. Solely reporting the accuracy on short windows in steady state can thus misrepresent the true temporal resolution that the algorithm achieves.

    Finally, the high conditional self-dependence between consecutive decisions for direct-classification algorithms also explains why their accuracy does not tend to improve with increasing window length, as shown in Figure \ref{fig:accWL} and in \cite{Geirnaert2020CSP, geirnaertRiemannianGeometryBasedDecoding2021a, Zhang, suSTAnetSpatiotemporalAttention2022, yanDARNetDualAttention2024}. In linear models, using longer windows is mathematically similar to averaging many shorter windows. When predictions are independent, averaging them leads to a reduction in variance, and thus an improved accuracy. However, when the predictions are conditionally dependent, the variance decreases at a much slower rate. 

\subsection{The accuracy and AAD scores around a switch}
    \label{subse:MI-accuracyInTrial}
\begin{figure*}[t]
    \centering
    \begin{subfigure}[t]{0.48\textwidth}
        \centering
        \includegraphics[width=\linewidth]{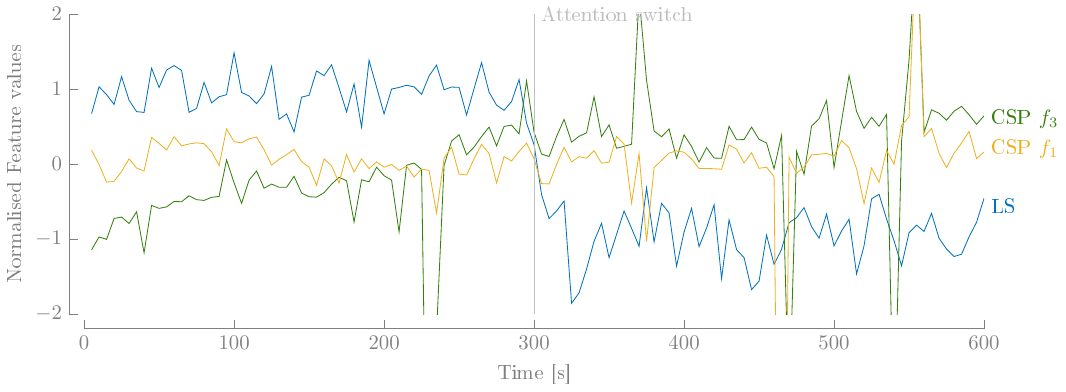}
        \caption{The average evolution of LS scores and the first and third CSP features after normalisation.}
    \end{subfigure}
    \hfill
    \begin{subfigure}[t]{0.48\textwidth}
        \centering
        \includegraphics[width=\linewidth]{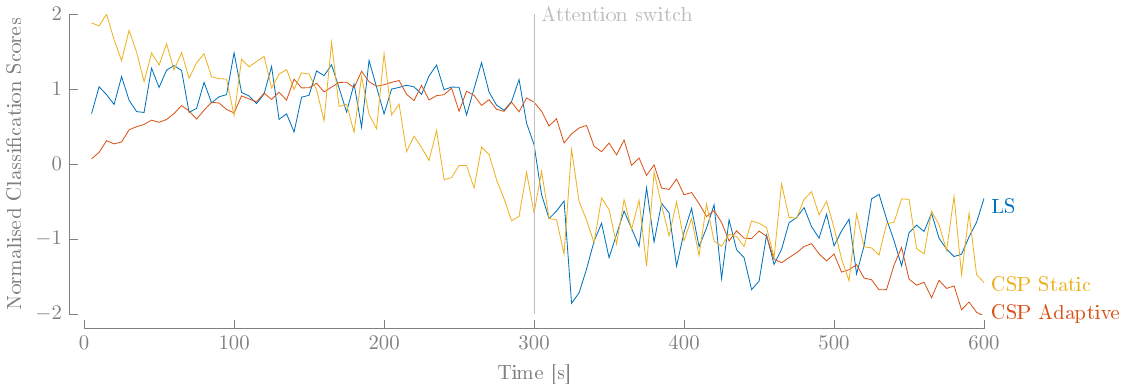}
        \caption{The average evolution of LS and CSP classification scores with static and adaptive LDA classifier after normalisation.}
        \label{fig:MI-classificationScoresAtSwitch}
    \end{subfigure}
    \caption{The average evolution of LS and CSP features and classification scores across a trial (for LS, we use the correlation coefficient both as the feature and classification score). At the attention switch, there is a clear jump in feature values for stimulus reconstruction algorithms such as LS. This is absent for direct-classification algorithms such as CSP. Instead, there is only a slow drift that can be misused by the classifier to obtain accuracies above chance level, despite the absence of any attention-driven effect.}
    \label{fig:MI-featAtSwitch}
\end{figure*}

    Figure \ref{fig:MI-featAtSwitch} shows the average evolution of the features and classification scores before and after an attention switch. The models were trained such that, for a perfect model, the features and classification scores should be high at the start, and immediately drop after the attention switch\footnote{All models were trained such that a feature value and classification score should be maximised for the left speaker, and minimised for the right speaker. For trials where the attention switches from the right speaker to the left speaker, the feature values and scores are multiplied by -1. This ensures that for every trial, the features and scores should start high, and drop to low values after the switch. The features and classification scores were then normalised such that that the average is 0 and the standard deviation is 1 for each model.}. This is, on average, the case for stimulus reconstruction models such as LS: there is a clear, sudden drop after the switch. However, the drop is not instant. This can be expected: we only know when the attention switch was cued. It is normal for a participant to take several seconds before they completed the attention switch, and it can be expected that the brain takes a few seconds to adapt to the new direction of attention. 

    However, there is no clear effect of an attention switch on the feature values of CSP. Some features (such as $f_1$) display inconsistent changes across trials (resulting in a flat slope on average), while other features (such as $f_3$) display clear slopes that are consistent across trials. Although we cannot say with certainty what causes these slopes, we hypothesise they are caused by temporal effects such as tiredness, or the aging of the electrode electrolyte, which causes the $\beta$-band filtered CSP features \cite{Geirnaert2020CSP} to change over time. However, it is certain that the slopes are not influenced by attention, as the attention switch has no effect on the direction or slope of the drift. 

    Depending on the classifier, these feature drifts can be abused in different ways, as shown in Figure \ref{fig:MI-classificationScoresAtSwitch}. The adaptive LDA classifier has access to the (oracle) pseudo-labels of the most recent CSP features. Due to the feature drifts, a new feature vector will lay most closely to the most recent features \cite{rotaruWhatAreWe2024}, and thus be assigned to the same class. Since attention switches are rare, this is most often correct, which allows the model to obtain a high accuracy, even if the features are in no way influenced by attention. Indeed, immediately after the attention switch, most feature vectors are still assigned to the first class. It is only after a sufficient amount of the most recent feature vectors in the training set belong to the second class, that new test features are also assigned to that class, leading to a high accuracy but a slow reaction to attention switches. The CSP model with adaptive LDA thus abuses the slow drifts in the features (which are not attention-related) to merely parrot the pseudo-labels provided to the LDA classifier. Since the pseudo-labels are in this case perfect (oracle labels), the accuracy is very high. But the only new information in the scores is driven by 'improving' the adaptive LDA such that it better reflects the feature drift within the trial. This causes the low, but non-zero rIMI. 

    Remarkably, while LOTO cross-validation is often considered to be insensitive to overfitting effects due to within-trial feature drift \cite{rotaruWhatAreWe2024, yanOverestimatedPerformanceAuditory2025, ivucicImpactCrossValidationSchemes2024}, we observe in Figure \ref{fig:MI-classificationScoresAtSwitch} that the static CSP scores also show a trend that seems to partially overlap with the change in attention (although the trend starts way before the switch such that it can not be attributed to attention)., and vice versa. Although this is far from consistent across trials, it suffices to obtain an accuracy slightly above chance level (see Figure \ref{fig:AccRimiWL}), despite its near-zero rIMI and the absence of any notable effect of attention on the features.     
    
    This demonstrates how even the accuracy obtained with LOTO cross-validation can be driven by a subtle form of information leakage from train to test set. With sufficiently complex models, such leakage can be exploited as soon as there is any predictable pattern in class labels across trials (which is typically the case because datasets are designed to be balanced across trials). 
    
    \subsection{The origin of the dependencies}
    \begin{figure}
        \centering
        \includegraphics[width=\linewidth]{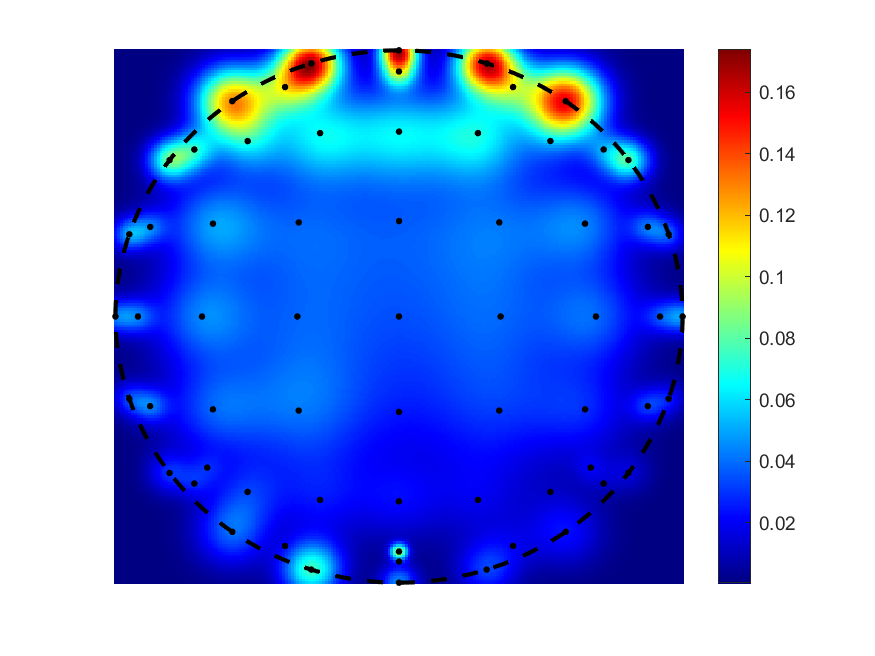}
        \caption{The distribution of the conditional self-dependency (CSD) of EEG across channels. More frontal channels display a significantly higher CSD.}
        \label{fig:topoPlot}
    \end{figure}
    In this last experiment, we wish to investigate why spatial and direct-classification AAD algorithms have an increased conditional dependency, which historically has led to their overestimated performances. This is done by investigating their raw inputs; the  per-window autocorrelation matrix $R_{mm}(n)$ for spatial and direct-classification AAD algorithms and crosscorrelation matrix $R_{my}(n)$ for stimulus reconstruction algorithms.
    
	On average, the conditional dependency $I(R_{mm}(i,j,n);R_{mm}(i,j,n-1)|y(n))$ of the individual elements of the sample autocorrelation matrix $R_{mm}(n)$ is \SI{0.04}{\bit}. For the sample crosscorrelation matrix $R_{my}(n)$, the conditional dependency $I(R_{my}(i,j,n);R_{my}(i,j,n-1)|y(n))$ is on average \SI{0.02}{\bit}. This confirms that there is (on average) more temporal dependency in the statistics used by spatial algorithms (the sample autocorrelation matrix) compared to stimulus reconstruction algorithms. 
	
	However, the distinctions become more apparent when the individual CSDs $I(R_{mm}(i,j,n);R_{mm}(i,j,n-1)|y(n))$ and $I(R_{my}(i,j,n);R_{my}(i,j,n-1)|y(n))$ are inspected in closer detail. In the crosscorrelation matrix, the dependencies are roughly equally distributed, and vary between \SIrange{0.01}{0.03}{\bit}. On the other hand, there is a clear pattern in the autocorrelation matrix, as shown in Figure \ref{fig:topoPlot}: the more frontal a channel is located, the more its autocorrelation and crosscorrelations with other frontal channels are temporally dependent. The conditional self-dependency ranges between \SIrange{0.1}{0.2}{\bit} for all frontal channels, an order of magnitude larger than the conditional dependency present in other channels. If these frontal channels are excluded, the average conditional dependency $I(R_{mm}(i,j,n);R_{mm}(i,j,n-1)|y(n)),\ i,j \not \in {F}$ drops to \SI{0.025}{\bit}, with $F$ containing the channels with prefix 'F'.
	
	This suggests that the majority of the conditional dependencies present in the features from spatial AAD algorithms stem from slow-moving processes in the frontal lobe. While eye gaze is one potential source of these dependencies, such long-range temporal correlations (LRTC) also occur when eye gaze is controlled, or eyes are closed, and are thus likely driven by complex neural interactions \cite{linkenkaer-hansenLongRangeTemporalCorrelations2001}. These dependencies remain significant for up to \SI{200}{\second}, and thus enables information leakage across trials when they are recorded in quick succession. 
	
	Because the CSD of consecutive predictions in CSP is caused by the underlying CSD of the sample autocorrelation matrix $R_{mm}(n)$, the overestimated performances obtained by ignoring CSD are unlikely to be limited to the algorithms discussed in this paper. Instead, it is expected to affect all or almost all spatial or other direct-classification AAD algorithms. While this hypothesis should be rigorously verified in future work, there are clear signs that this phenomenon is indeed inherent to the broader framework of spatial and direct-classification algorithms. For instance, the reported accuracies of such direct-classification AAD algorithms typically barely increase for longer window lengths \cite{Geirnaert2020a,Geirnaert2020CSP,geirnaertRiemannianGeometryBasedDecoding2021a,suSTAnetSpatiotemporalAttention2022,Ciccarelli2019,vandecappelleEEGbasedDetectionLocus2021,khanShortwindowEEGbasedAuditory2025}. They also consistently drop in performance when tested on LOTO CV \cite{rotaruWhatAreWe2024, ivucicImpactCrossValidationSchemes2024, yanOverestimatedPerformanceAuditory2025}. As clarified previously, this strongly indicates that the performance is largely driven by irrelevant feature drift (akin to Brownian motion), rather than an attention-driven component. In contrast, the accuracy of stimulus reconstruction algorithms consistently increases with increasing window length, no matter whether the stimulus was reconstructed using linear or non-linear models \cite{OSullivan2017, DeCheveigne2018, nguyenAADNetEndtoEndDeep2025,sridharImprovingAuditoryAttention2025}. Note that hybrid AAD algorithms, which directly classify a combination of EEG and envelopes by minimising classification loss, also behave like a direct-classification model, and are equally affected by these dependencies. An example of such a hybrid model is the model proposed by Cicarelli et al. \cite{Ciccarelli2019}. Similar to EEG-based direct-classification algorithms, the reported accuracies of these models barely increase for longer window lengths.

    Finally, although the high conditional dependencies in EEG are an important driver of the conditional dependencies in spatial AAD predictions, they are not necessarily the only driver. Various model architectures, such as LSTM's, state space machines and recurrent neural networks can implicitly include a 'memory' component, which also create conditional dependencies across consecutive predictions. Although these dependencies are often desired, it is important to compare such algorithms fairly to AAD algorithms where such components are not (yet) added, and to take note of both the steady state performance and the temporal resolution around an attention switch, or use the rIMI metric when insufficient attention switches are available. 
	
	\section{Conclusion}
    AAD algorithms are currently (almost) uniquely validated by measuring their accuracy or ITR in function of the window length during a steady state, where a listener continuously listens to the same speaker without or with limited intra-trial attention switches. However, such metrics can be misleading, as they implicitly assume conditional independence between consecutive AAD predictions. This assumption is often violated, especially by direct-classification AAD algorithms that directly classify the EEG without explicitly relating it to the speech signal. When these conditional dependencies are not accounted for, the true temporal resolution of an algorithm can be severely overestimated. 

    To correct for these dependencies, we have proposed a new metric called relative Incremental Mutual Information (rIMI), which measures what fraction of the remaining uncertainty about the attention state a new prediction removes, given all previous predictions. This metric rewards predictions that are both accurate and provide new information not previously available. It demonstrates that direct-classification AAD algorithms remove uncertainty at a much lower rate than stimulus reconstruction algorithms in steady state. By also investigating the behaviour of these AAD models around an attention switch, we have discovered that the higher apparent performance of direct-classification AAD higher accuracy is a mirage obtained by exploiting temporal drift, which leaks information from training to test data. This information leakage is even possible when leave-one-out cross-validation is used, a cross-validation scheme that is generally assumed to avoid inflated accuracies due to overfitting to feature drift. 

    The conditional self-dependence present in direct-classification algorithms is caused by Brownian-like drift of the EEG statistics, which is naturally present in EEG recordings. The overestimated performances reported in this paper can thus affect any EEG decoding paradigm where the decoded states rarely change.  
    
    \appendices
    \section{The adaptive CSP algorithm}
    \label{app:adaptive CSP algorithm}
    The adaptive CSP algorithm uses the same CSP filters as the static CSP algorithm \cite{Geirnaert2020CSP}, which is trained in a supervised fashion using leave-one-trial-out cross-validation. However, the LDA classifier is adaptively retrained with oracle labels. The adaptive algorithm thus merely serves as an upper bound of what a practical unsupervised adaptive CSP algorithm could achieve. Given a CSP feature vector $\Vec{x}(n)$ that belongs to either to class $\mathcal{C}_+$ or $\mathcal{C}_-$, we update the estimated class averages $\bm{\mu}_\pm$, class covariances $\Sigma_\pm$, LDA decoder $\Vec{d}$ and threshold $\theta$ as follows:
    
    \begin{align*} 
    &\text{if } n \in \mathcal{C}_+:\\ 
    &\quad \bm{\mu}_+ \xleftarrow{} \lambda \bm{\mu}_+ + (1-\lambda)\Vec{x}(n)\\
    & \quad \tilde{\Vec{x}}(n) = \Vec{x}(n) - \bm{\mu}_+\\
    &\quad \Sigma_+ \xleftarrow{} \lambda \Sigma_+ + (1-\lambda) \tilde{\Vec{x}}(n)\tilde{\Vec{x}}^\top(n)\\
    &\quad \bm{\mu}_- \xleftarrow{} \bm{\mu}_-\\
    &\quad \Sigma_- \xleftarrow{} \Sigma_-\\
    &\\
    & \text{else if } n \in \mathcal{C}_-:\\
       &\quad \bm{\mu}_- \xleftarrow{} \lambda \bm{\mu}_- + (1-\lambda)\Vec{x}(n)\\
    & \quad \tilde{\Vec{x}}(n) = \Vec{x}(n) - \bm{\mu}_-\\
    &\quad \Sigma_- \xleftarrow{} \lambda \Sigma_- + (1-\lambda) \tilde{\Vec{x}}(n)\tilde{\Vec{x}}^\top(n)\\
    &\quad \bm{\mu}_+ \xleftarrow{} \bm{\mu}_+\\
    &\quad \Sigma_+ \xleftarrow{} \Sigma_+\\
    &\\
    &\Vec{d} = (\Sigma_+ + \Sigma_-)^{-1}(\bm{\mu}_1 - \bm{\mu}_2)\\
    &\theta = \Vec{d}^\top (\bm{\mu}_1 + \bm{\mu}_2)/2
    \end{align*}

    The classification score $f(n+1)$ of the next window is then computed as: 
    \begin{equation}
        f(n+1) = \Vec{d}^\top\Vec{x}(n+1)-\theta.
    \end{equation}
    The class averages $\bm{\mu}_\pm$ and covariances $\Sigma_\pm$ are initialised using all left-out trials. The forgetting factor $\lambda$ is set to $\frac{N_{eff}/\tau-1}{N_{eff}/\tau+1}$, with $N_{eff}$ the equivalent sliding window length equal to \SI{600}{\second}, and $\tau$ the decision window length. 
    
    \bibliographystyle{IEEEtran}
	\bibliography{library}

@article{Biesmans2017,
  title = {Auditory-Inspired Speech Envelope Extraction Methods for Improved {{EEG-based}} Auditory Attention Detection in a Cocktail Party Scenario},
  author = {Biesmans, Wouter and Das, Neetha and Francart, Tom and Bertrand, Alexander},
  year = 2017,
  journal = {IEEE Transactions on Neural Systems and Rehabilitation Engineering},
  volume = {25},
  number = {5},
  pages = {402--412},
  issn = {15344320},
  doi = {10.1109/TNSRE.2016.2571900}
}

@article{Cherry1953,
  title = {Some {{Experiments}} on the {{Recognition}} of {{Speech}}, with {{One}} and with {{Two Ears}}.},
  author = {Cherry, E. Colin},
  year = 1953,
  journal = {Journal of the Acoustical Society of America},
  volume = {25},
  pages = {974--979},
  doi = {http://dx.doi.org/10.1121/1.1907229}
}

@article{Ciccarelli2019,
  title = {Comparison of {{Two-Talker Attention Decoding}} from {{EEG}} with {{Nonlinear Neural Networks}} and {{Linear Methods}}},
  author = {Ciccarelli, Gregory and Nolan, Michael and Perricone, Joseph and Calamia, Paul T. and Haro, Stephanie and O'Sullivan, James and Mesgarani, Nima and Quatieri, Thomas F. and Smalt, Christopher J.},
  year = 2019,
  journal = {Scientific Reports},
  volume = {9},
  issn = {2045-2322},
  doi = {10.1038/s41598-019-47795-0}
}

@article{DeCheveigne2018,
  title = {Decoding the Auditory Brain with Canonical Component Analysis},
  author = {{de Cheveign{\'e}}, Alain and Wong, Daniel E. and Di Liberto, Giovanni M. and Hjortkj{\ae}r, Jens and Slaney, Malcolm and Lalor, Edmund},
  year = 2018,
  journal = {NeuroImage},
  volume = {172},
  pages = {206--216},
  issn = {10959572},
  doi = {10.1016/j.neuroimage.2018.01.033},
  urldate = {2020-04-27}
}

@article{Geirnaert2020,
  title = {An {{Interpretable Performance Metric}} for {{Auditory Attention Decoding Algorithms}} in a {{Context}} of {{Neuro-Steered Gain Control}}},
  author = {Geirnaert, Simon and Francart, Tom and Bertrand, Alexander},
  year = 2020,
  journal = {IEEE Transactions on Neural Systems and Rehabilitation Engineering},
  volume = {28},
  number = {1},
  pages = {307--317},
  issn = {15580210},
  doi = {10.1109/TNSRE.2019.2952724}
}

@article{Geirnaert2020a,
  title = {Electroencephalography-{{Based Auditory Attention Decoding}}: {{Toward Neurosteered Hearing Devices}}},
  author = {Geirnaert, Simon and Vandecappelle, Servaas and Alickovic, Emina and {de Cheveign{\'e}}, Alain and Lalor, Edmund and Meyer, Bernd T. and Miran, Sina and Francart, Tom and Bertrand, Alexander},
  year = 2021,
  journal = {IEEE Signal Processing Magazine},
  volume = {38},
  number = {4},
  pages = {89--102},
  doi = {10.1109/MSP.2021.3075932},
  urldate = {2020-09-03}
}

@article{Geirnaert2020CSP,
  title = {Fast {{EEG-based}} Decoding of the Directional Focus of Auditory Attention Using Common Spatial Patterns},
  author = {Geirnaert, Simon and Francart, Tom and Bertrand, Alexander},
  year = 2020,
  journal = {IEEE Transactions on Biomedical Engineering},
  doi = {10.1109/TBME.2020.3033446},
  urldate = {2020-09-01},
  pmid = {33095706}
}

@article{Geirnaert2021Unsup,
  title = {Unsupervised {{Self-Adaptive Auditory Attention Decoding}}},
  author = {Geirnaert, Simon and Francart, Tom and Bertrand, Alexander},
  year = 2021,
  journal = {IEEE Journal of Biomedical and Health Informatics},
  volume = {25},
  number = {10},
  pages = {3955--3966},
  issn = {2168-2194},
  doi = {10.1109/JBHI.2021.3075631},
  urldate = {2021-05-04}
}

@article{Geirnaert2022Unsup,
  title = {Time-Adaptive {{Unsupervised Auditory Attention Decoding Using EEG-based Stimulus Reconstruction}}},
  author = {Geirnaert, Simon and Francart, Tom and Bertrand, Alexander},
  year = 2022,
  journal = {IEEE Journal of Biomedical and Health Informatics},
  volume = {26},
  number = {8},
  doi = {10.1109/JBHI.2022.3162760},
  urldate = {2022-04-13}
}

@inproceedings{geirnaertRiemannianGeometryBasedDecoding2021a,
  title = {Riemannian {{Geometry-Based Decoding}} of the {{Directional Focus}} of {{Auditory Attention Using EEG}}},
  booktitle = {{{ICASSP}} 2021 - 2021 {{IEEE International Conference}} on {{Acoustics}}, {{Speech}} and {{Signal Processing}} ({{ICASSP}})},
  author = {Geirnaert, Simon and Francart, Tom and Bertrand, Alexander},
  year = 2021,
  month = jun,
  pages = {1115--1119},
  issn = {2379-190X},
  doi = {10.1109/ICASSP39728.2021.9413404},
  urldate = {2026-08-12}
}

@inproceedings{Heintz2023Unbiased,
  title = {Unbiased {{Unsupervised Stimulus Reconstruction}} for {{EEG-Based Auditory Attention Decoding}}},
  booktitle = {2023 {{IEEE International Conference}} on {{Acoustics}}, {{Speech}} and {{Signal Processing}} ({{ICASSP}})},
  author = {Heintz, Nicolas and Geirnaert, Simon and Francart, Tom and Bertrand, Alexander},
  year = 2023,
  publisher = {{Institute of Electrical and Electronics Engineers (IEEE)}},
  doi = {10.1109/ICASSP49357.2023.10096608},
  urldate = {2023-10-16},
  copyright = {All rights reserved}
}

@article{heintzPostprocessingEEGbasedAuditory2025,
  title = {Post-Processing of {{EEG-based Auditory Attention Decoding Decisions}} via {{Hidden Markov Models}}},
  author = {Heintz, Nicolas and Francart, Tom and Bertrand, Alexander},
  year = 2025,
  month = jun,
  journal = {arXiv preprint arXiv:2506.24024},
  eprint = {2506.24024},
  primaryclass = {eess},
  doi = {10.48550/arXiv.2506.24024},
  urldate = {2025-09-24},
  archiveprefix = {arXiv}
}

@inproceedings{heintzProbabilisticGainControl2024,
  title = {Probabilistic {{Gain Control}} in a {{Multi-Speaker Setting Using EEG-Based Auditory Attention Decoding}}},
  booktitle = {2024 32nd {{European Signal Processing Conference}} ({{EUSIPCO}})},
  author = {Heintz, Nicolas and Geirnaert, Simon and {Van de Ryck}, Iris and Francart, Tom and Bertrand, Alexander},
  year = 2024,
  month = aug,
  pages = {892--896},
  issn = {2076-1465},
  doi = {10.23919/EUSIPCO63174.2024.10715360},
  urldate = {2025-02-19}
}

@article{heintzUnsupervisedEEGbasedDecoding2025,
  title = {Unsupervised {{EEG-based}} Decoding of Absolute Auditory Attention with Canonical Correlation Analysis},
  author = {Heintz, Nicolas and Francart, Tom and Bertrand, Alexander},
  year = 2025,
  month = apr,
  journal = {arXiv preprint arXiv:2504.17724},
  eprint = {2504.17724},
  primaryclass = {eess},
  doi = {10.48550/arXiv.2504.17724},
  urldate = {2025-09-24},
  archiveprefix = {arXiv}
}

@article{hjortkjaerRealtimeControlHearing2025,
  title = {Real-Time Control of a Hearing Instrument with {{EEG-based}} Attention Decoding},
  author = {Hjortkj{\ae}r, Jens and Wong, Daniel D E and Catania, Alessandro and {M{\"a}rcher-R{\o}rsted}, Jonatan and Ceolini, Enea and Fuglsang, S{\o}ren A and Kiselev, Ilya and Di Liberto, Giovanni and Liu, Shih-Chii and Dau, Torsten and Slaney, Malcolm and {de Cheveign{\'e}}, Alain},
  year = 2025,
  month = feb,
  journal = {Journal of Neural Engineering},
  volume = {22},
  number = {1},
  pages = {016027},
  publisher = {IOP Publishing},
  issn = {1741-2552},
  doi = {10.1088/1741-2552/ad867c},
  urldate = {2025-08-27},
  langid = {english}
}

@inproceedings{huberEntropyApproximationGaussian2008,
  title = {On Entropy Approximation for {{Gaussian}} Mixture Random Vectors},
  booktitle = {2008 {{IEEE International Conference}} on {{Multisensor Fusion}} and {{Integration}} for {{Intelligent Systems}}},
  author = {Huber, Marco F. and Bailey, Tim and {Durrant-Whyte}, Hugh and Hanebeck, Uwe D.},
  year = 2008,
  month = aug,
  pages = {181--188},
  publisher = {IEEE},
  address = {Seoul, Korea (South)},
  doi = {10.1109/MFI.2008.4648062},
  urldate = {2025-09-20},
  copyright = {https://doi.org/10.15223/policy-029},
  isbn = {978-1-4244-2143-5}
}

@article{inceStatisticalFrameworkNeuroimaging2017,
  title = {A Statistical Framework for Neuroimaging Data Analysis Based on Mutual Information Estimated via a Gaussian Copula},
  author = {Ince, Robin A.A. and Giordano, Bruno L. and Kayser, Christoph and Rousselet, Guillaume A. and Gross, Joachim and Schyns, Philippe G.},
  year = 2017,
  journal = {Human Brain Mapping},
  volume = {38},
  number = {3},
  pages = {1541--1573},
  issn = {1097-0193},
  doi = {10.1002/hbm.23471},
  urldate = {2025-09-19},
  langid = {english}
}

@inproceedings{ivucicImpactCrossValidationSchemes2024,
  title = {The {{Impact}} of {{Cross-Validation Schemes}} for {{EEG-Based Auditory Attention Detection}} with {{Deep Neural Networks}}},
  booktitle = {2024 46th {{Annual International Conference}} of the {{IEEE Engineering}} in {{Medicine}} and {{Biology Society}} ({{EMBC}})},
  author = {Ivucic, Gabriel and Pahuja, Saurav and Putze, Felix and Cai, Siqi and Li, Haizhou and Schultz, Tanja},
  year = 2024,
  month = jul,
  pages = {1--4},
  issn = {2694-0604},
  doi = {10.1109/EMBC53108.2024.10782636},
  urldate = {2025-05-26}
}

@article{khanShortwindowEEGbasedAuditory2025,
  title = {Short-Window {{EEG-based}} Auditory Attention Decoding for Neuroadaptive Hearing Support for Smart Healthcare},
  author = {Khan, Ihtiram Raza and Peng, Sheng-Lung and Mahajan, Rupali and Dey, Rajesh},
  year = 2025,
  month = sep,
  journal = {Neuroscience Informatics},
  volume = {5},
  number = {3},
  pages = {100222},
  issn = {2772-5286},
  doi = {10.1016/j.neuri.2025.100222},
  urldate = {2025-09-21}
}

@article{linkenkaer-hansenLongRangeTemporalCorrelations2001,
  title = {Long-{{Range Temporal Correlations}} and {{Scaling Behavior}} in {{Human Brain Oscillations}}},
  author = {{Linkenkaer-Hansen}, Klaus and Nikouline, Vadim V. and Palva, J. Matias and Ilmoniemi, Risto J.},
  year = 2001,
  month = feb,
  journal = {The Journal of Neuroscience},
  volume = {21},
  number = {4},
  pages = {1370--1377},
  issn = {0270-6474},
  doi = {10.1523/JNEUROSCI.21-04-01370.2001},
  urldate = {2026-08-11},
  pmcid = {PMC6762238},
  pmid = {11160408}
}

@article{nguyenAADNetEndtoEndDeep2025,
  title = {{{AADNet}}: {{An End-to-End Deep Learning Model}} for {{Auditory Attention Decoding}}},
  shorttitle = {{{AADNet}}},
  author = {Nguyen, Nhan Duc Thanh and Phan, Huy and Geirnaert, Simon and Mikkelsen, Kaare and Kidmose, Preben},
  year = 2025,
  journal = {IEEE Transactions on Neural Systems and Rehabilitation Engineering},
  volume = {33},
  pages = {2695--2706},
  issn = {1558-0210},
  doi = {10.1109/TNSRE.2025.3587637},
  urldate = {2026-08-12}
}

@article{OSullivan2014,
  title = {Attentional {{Selection}} in a {{Cocktail Party Environment Can Be Decoded}} from {{Single-Trial EEG}}},
  author = {O'Sullivan, James A. and Power, Alan J. and Mesgarani, Nima and Rajaram, Siddharth and Foxe, John J. and {Shinn-Cunningham}, Barbara G. and Slaney, Malcolm and Shamma, Shihab A. and Lalor, Edmund C.},
  year = 2015,
  journal = {Cerebral Cortex},
  volume = {25},
  number = {7},
  pages = {1697--1706},
  issn = {14602199},
  doi = {10.1093/cercor/bht355},
  pmid = {24429136}
}

@article{OSullivan2017,
  title = {Neural Decoding of Attentional Selection in Multi-Speaker Environments without Access to Clean Sources.},
  author = {O'Sullivan, James and Chen, Zhuo and Herrero, Jose and McKhann, Guy M and Sheth, Sameer A and Mehta, Ashesh D and Mesgarani, Nima},
  year = 2017,
  journal = {Journal of neural engineering},
  volume = {14},
  number = {5},
  pages = {056001},
  issn = {1741-2552},
  doi = {10.1088/1741-2552/aa7ab4},
  urldate = {2020-05-01},
  pmid = {28776506}
}

@inproceedings{Presacco2019,
  title = {Real-{{Time Tracking}} of {{Magnetoencephalographic Neuromarkers}} during a {{Dynamic Attention-Switching Task}}},
  booktitle = {Proceedings of the {{Annual International Conference}} of the {{IEEE Engineering}} in {{Medicine}} and {{Biology Society}}, {{EMBS}}},
  author = {Presacco, Alessandro and Miran, Sina and Babadi, Behtash and Simon, Jonathan Z.},
  year = 2019,
  month = jul,
  pages = {4148--4151},
  publisher = {{Institute of Electrical and Electronics Engineers Inc.}},
  doi = {10.1109/EMBC.2019.8857953},
  urldate = {2020-05-01},
  isbn = {978-1-5386-1311-5}
}

@article{presaccoSpeechinnoiseRepresentationAging2019,
  title = {Speech-in-Noise Representation in the Aging Midbrain and Cortex: {{Effects}} of Hearing Loss},
  shorttitle = {Speech-in-Noise Representation in the Aging Midbrain and Cortex},
  author = {Presacco, Alessandro and Simon, Jonathan Z. and Anderson, Samira},
  year = 2019,
  month = mar,
  journal = {PLOS ONE},
  volume = {14},
  number = {3},
  pages = {e0213899},
  publisher = {Public Library of Science},
  issn = {1932-6203},
  doi = {10.1371/journal.pone.0213899},
  urldate = {2025-08-29},
  langid = {english}
}

@misc{rotaruAudiovisualGazecontrolledAuditory2024,
  title = {Audiovisual, {{Gaze-controlled Auditory Attention Decoding Dataset KU Leuven}} ({{AV-GC-AAD}})},
  author = {Rotaru, Iustina and Geirnaert, Simon and Bertrand, Alexander and Francart, Tom},
  year = 2024,
  month = apr,
  publisher = {Zenodo},
  urldate = {2025-06-24}
}

@article{rotaruWhatAreWe2024,
  title = {What Are We Really Decoding? {{Unveiling}} Biases in {{EEG-based}} Decoding of the Spatial Focus of Auditory Attention},
  shorttitle = {What Are We Really Decoding?},
  author = {Rotaru, Iustina and Geirnaert, Simon and Heintz, Nicolas and {Van de Ryck}, Iris and Bertrand, Alexander and Francart, Tom},
  year = 2024,
  month = feb,
  journal = {Journal of Neural Engineering},
  volume = {21},
  number = {1},
  pages = {016017},
  publisher = {IOP Publishing},
  issn = {1741-2552},
  doi = {10.1088/1741-2552/ad2214},
  urldate = {2025-05-13},
  langid = {english}
}

@article{Shinn-Cunningham2008,
  title = {Selective {{Attention}} in {{Normal}} and {{Impaired Hearing}}},
  author = {{Shinn-Cunningham}, Barbara G. and Best, Virginia},
  year = 2008,
  journal = {Trends in Amplification},
  volume = {12},
  number = {4},
  pages = {283--299},
  publisher = {SAGE Publications},
  issn = {19405588},
  doi = {10.1177/1084713808325306},
  urldate = {2020-04-27}
}

@article{sridharImprovingAuditoryAttention2025,
  title = {Improving Auditory Attention Decoding in Noisy Environments for Listeners with Hearing Impairment through Contrastive Learning},
  author = {Sridhar, Gautam and Boselli, Sof{\'i}a and Skoglund, Martin A and Bernhardsson, Bo and Alickovic, Emina},
  year = 2025,
  month = jun,
  journal = {Journal of Neural Engineering},
  volume = {22},
  number = {3},
  pages = {036041},
  publisher = {IOP Publishing},
  issn = {1741-2552},
  doi = {10.1088/1741-2552/ade28a},
  urldate = {2025-09-17},
  langid = {english}
}

@article{suSTAnetSpatiotemporalAttention2022,
  title = {{{STAnet}}: {{A Spatiotemporal Attention Network}} for {{Decoding Auditory Spatial Attention From EEG}}},
  shorttitle = {{{STAnet}}},
  author = {Su, Enze and Cai, Siqi and Xie, Longhan and Li, Haizhou and Schultz, Tanja},
  year = 2022,
  month = jul,
  journal = {IEEE Transactions on Biomedical Engineering},
  volume = {69},
  number = {7},
  pages = {2233--2242},
  issn = {1558-2531},
  doi = {10.1109/TBME.2022.3140246},
  urldate = {2025-05-26}
}

@article{vandecappelleEEGbasedDetectionLocus2021,
  title = {{{EEG-based}} Detection of the Locus of Auditory Attention with Convolutional Neural Networks},
  author = {Vandecappelle, Servaas and Deckers, Lucas and Das, Neetha and Ansari, Amir Hossein and Bertrand, Alexander and Francart, Tom},
  editor = {{Shinn-Cunningham}, Barbara G and O'Sullivan, James and Dimitrijevic, Andrew},
  year = 2021,
  month = apr,
  journal = {eLife},
  volume = {10},
  pages = {e56481},
  publisher = {eLife Sciences Publications, Ltd},
  issn = {2050-084X},
  doi = {10.7554/eLife.56481},
  urldate = {2025-09-30}
}

@article{xuBewareOverestimatedDecoding2024,
  title = {Beware of {{Overestimated Decoding Performance Arising}} from {{Temporal Autocorrelations}} in {{Electroencephalogram Signals}}},
  author = {Xu, Xiran and Wang, Bo and Xiao, Boda and Niu, Yadong and Wang, Yiwen and Wu, Xihong and Chen, Jing},
  year = 2024,
  month = may,
  journal = {arXiv preprint arXiv:.2405.17024},
  eprint = {2405.17024},
  primaryclass = {eess},
  doi = {10.48550/arXiv.2405.17024},
  urldate = {2025-09-30},
  archiveprefix = {arXiv}
}

@article{yanDARNetDualAttention2024,
  title = {{{DARNet}}: {{Dual Attention Refinement Network}} with {{Spatiotemporal Construction}} for {{Auditory Attention Detection}}},
  shorttitle = {{{DARNet}}},
  author = {Yan, Sheng and {fan}, Cunhang and Zhang, Hongyu and Yang, Xiaoke and Tao, Jianhua and Lv, Zhao},
  year = 2024,
  month = nov,
  journal = {arXiv preprint arXiv:2410.11181},
  eprint = {2410.11181},
  primaryclass = {eess},
  doi = {10.48550/arXiv.2410.11181},
  urldate = {2025-12-23},
  archiveprefix = {arXiv}
}

@inproceedings{yanOverestimatedPerformanceAuditory2025,
  title = {Overestimated Performance of Auditory Attention Decoding Caused by Experimental Design in {{EEG}} Recordings},
  booktitle = {Proc. {{Interspeech}} 2025},
  author = {Yan, Yujie and Xu, Xiran and Zhu, Haolin and Li, Songyi and Wang, Bo and Wu, Xihong and Chen, Jing},
  year = 2025,
  pages = {1053--1057},
  doi = {10.21437/Interspeech.2025-85},
  urldate = {2025-08-25},
  langid = {english}
}

@inproceedings{Zhang,
  title = {{{EEG-based Short-time Auditory Attention Detection}} Using {{Multi-task Deep Learning}}},
  booktitle = {Interspeech 2020},
  author = {Zhang, Zhuo and Zhang, Gaoyan and Dang, Jianwu and Wu, Shuang and Zhou, Di and Wang, Longbiao},
  year = 2020,
  pages = {2517--2521},
  doi = {10.21437},
  urldate = {2020-10-23}
}
\end{document}